\documentclass[smallextended]{svjour3}       
\smartqed  
\usepackage{amsmath}
\usepackage{graphicx}
\usepackage[utf8]{inputenc}

\usepackage[hyphens]{url} 
\usepackage{hyperref}

\usepackage{amsmath} \usepackage{subfig} \usepackage{float} 

\usepackage{color}
\usepackage{fancyvrb}

\DefineVerbatimEnvironment{Highlighting}{Verbatim}{commandchars=\\\{\}}
\usepackage{framed}
\definecolor{shadecolor}{RGB}{248,248,248}
\newenvironment{Shaded}{\begin{snugshade}}{\end{snugshade}}

\newcommand{\CommentTok}[1]{\textcolor[rgb]{0.56,0.35,0.01}{\textit{#1}}}

\newcommand{\ControlFlowTok}[1]{\textcolor[rgb]{0.13,0.29,0.53}{\textbf{#1}}}
\newcommand{\DataTypeTok}[1]{\textcolor[rgb]{0.13,0.29,0.53}{#1}}
\newcommand{\DecValTok}[1]{\textcolor[rgb]{0.00,0.00,0.81}{#1}}

\newcommand{\FloatTok}[1]{\textcolor[rgb]{0.00,0.00,0.81}{#1}}

\newcommand{\KeywordTok}[1]{\textcolor[rgb]{0.13,0.29,0.53}{\textbf{#1}}}
\newcommand{\NormalTok}[1]{#1}
\newcommand{\OperatorTok}[1]{\textcolor[rgb]{0.81,0.36,0.00}{\textbf{#1}}}
\newcommand{\OtherTok}[1]{\textcolor[rgb]{0.56,0.35,0.01}{#1}}

\newcommand{\StringTok}[1]{\textcolor[rgb]{0.31,0.60,0.02}{#1}}

\usepackage{float}
\let\origfigure\figure
\let\endorigfigure\endfigure
\renewenvironment{figure}[1][2] {
    \expandafter\origfigure\expandafter[H]
} {
    \endorigfigure
}

\newcommand\independent{\protect\mathpalette{\protect\independenT}{\perp}}
\def\independenT#1#2{\mathrel{\rlap{$#1#2$}\mkern2mu{#1#2}}}

\begin{document}

\title{ROC and AUC with a Binary Predictor: a Potentially Misleading Metric \thanks{This analysis was supported by NIH Grants R01NS060910 and U01NS080824.} }

    \titlerunning{Binary Predictor ROC}

\author{ John Muschelli }

\date{\href{mailto:jmusche1@jhu.edu}{jmusche1@jhu.edu} \\
      Department of Biostatistics, \\
      Johns Hopkins Bloomberg School of Public Health \\
      615 N Wolfe St \\
      Baltimore, MD 21205 \\
      }

\maketitle

\begin{abstract}
In analysis of binary outcomes, the receiver operator characteristic
(ROC) curve is heavily used to show the performance of a model or
algorithm. The ROC curve is informative about the performance over a
series of thresholds and can be summarized by the area under the curve
(AUC), a single number. When a \textbf{predictor} is categorical, the
ROC curve has one less than number of categories as potential
thresholds; when the predictor is binary there is only one threshold. As
the AUC may be used in decision-making processes on determining the best
model, it important to discuss how it agrees with the intuition from the
ROC curve. We discuss how the interpolation of the curve between
thresholds with binary predictors can largely change the AUC. Overall,
we show using a linear interpolation from the ROC curve with binary
predictors corresponds to the estimated AUC, which is most commonly done
in software, which we believe can lead to misleading results. We compare
R, Python, Stata, and SAS software implementations. We recommend using
reporting the interpolation used and discuss the merit of using the step
function interpolator, also referred to as the ``pessimistic'' approach
by Fawcett (2006).
\\
\keywords{
        roc \and
        auc \and
        area under the curve \and
        R \and
    }

\end{abstract}

\def\spacingset#1{\renewcommand{\baselinestretch}%
{#1}\small\normalsize} \spacingset{1}

\let\code=\texttt
\let\proglang=\textsf
\newcommand{\pkg}[1]{{\fontseries{b}\selectfont #1}}

\hypertarget{introduction}{%
\section{Introduction}\label{introduction}}

In many applications, receiver operator characteristic (ROC) curves are
used to show how a predictor compares to the true outcome. One of the
large advantages of ROC analysis is that it is threshold-agnostic;
performance of a predictor is estimated without a specific threshold and
also gives a criteria to choose an optimal threshold based on a certain
cost function or objective. Typically, an ROC analysis shows how
sensitivity (true positive rate) changes with varying specificity (true
negative rate or \(1 - \text{false positive rate}\)) for different
thresholds. Analyses also typically weigh false positives and false
negatives equally. In ROC analyses, the predictive capabilities of a
variable is commonly summarized by the area under the curve (AUC), which
can be found by integrating areas under the line segments. We will
discuss how interpolation between these line segments affect the
visualization of the ROC curve and corresponding AUC. Additionally,
partial ROC (pROC) analysis keeps a maximum specificity fixed and can
summarize a predictor by the partial AUC (pAUC), integrating up to the
maximum specificity, or the maximum sensitivity with the smallest false
positive rate in that subset range.

Many predictors, especially medical tests, result in a binary decision;
a value is higher than a pre-determined threshold or a substance is
present. Similarly, some predictors are commonly collected as
categorical or discrete such as low, normal, or high blood pressure
while others are categorical by nature such as having a specific gene or
not. These are useful indicators of presence a disease, which is a
primary outcome of interest in medical settings, and are used heavily in
analysis.

If one assumes the binary predictor is generated from a continuous
distribution that has been thresholded, then the sensitivity of this
thresholded predictor actually represents one point on the ROC curve for
the underlying continuous value. Therefore the ROC curve of a binary
predictor is not really appropriate, but should be represented by a
single point on the curve. But alas, ROC and AUC analysis has been done
on binary predictors and used to inform if one variable is more
predictive than the other (E et al. 2018; TV et al. 2017; Glaveckaite et
al. 2011; Blumberg et al. 2016; Budwega et al. 2016; Mwipatayi et al.
2016; Xiong et al. 2018, @shterev2018bayesian; Kushnir et al. 2018;
Snarr et al. 2017; Veltri, Kamath, and Shehu 2018). For example, these
cases show that researchers use ROC curves and AUC to evaluate
predictors, even when the predictors are categorical or binary. Although
there is nothing inherently wrong with this comparison, it can lead to
drastically different predictors being selected based on these criteria
if ties are treated slightly different ways. A more appropriate
comparison of a continuous predictor and the binary predictor may be to
compare the sensitivity and specificity (or overall accuracy) of the
continuous predictor given the optimal threshold versus that of the
binary predictor.

As categorical/binary predictors only have a relatively small number of
categories, how ties are handled are distinctly relevant. Thus, many
observations may have the same value/risk score. Fawcett (2006)
describes the standard way of how ties are handled in a predictor: a
probability of \(\frac{1}{2}\) is given for the cases when the
predictors are tied. When drawing the ROC curve, one can assume that all
the ties do not correctly classify the outcome (Fawcett called the
``pessimistic'' approach) or that all the ties do correctly classify the
outcome (called the ``optimistic'' approach), see Fig. 6 in (Fawcett
2006). But Fawcett notes (emphasis in original):

\begin{quote}
Any mixed ordering of the instances will give a different set of step
segments within the rectangle formed by these two extremes. However, the
ROC curve should represent the \emph{expected} performance of the
classifier, which, lacking any other information, is the average of the
pessimistic and optimistic segments.
\end{quote}

This ``expected'' performance directly applies to the assignment of a
half probability of success when the data are tied, which is implied by
the ``trapezoidal rule'' from Hanley and McNeil (1982). Fawcett (2006)
also states in the calculation of AUC that ``trapezoids are used rather
than rectangles in order to average the effect between points''. This
trapezoidal rule applies additional areas to the AUC based on ties of
the predictor, giving a half probability. This addition of half
probability is linked to how ties are treated in the Wilcoxon rank sum
test. As much of the theory of ROC curve testing, and therefore testing
of differences in AUC, is based on the theory of the Wilcoxon rank-sum
test, this treatment of ties is also relevant to statistical inference
and not only AUC estimation.

Others have discussed insights into binary predictors in addition to
Fawcett (2006), but they are mentioned in small sections of the paper
(Saito and Rehmsmeier 2015; Pepe, Longton, and Janes 2009). Other
information regarding ties and binary data are blog posts or working
papers such as
\url{http://blog.revolutionanalytics.com/2016/11/calculating-auc.html}
or \url{https://www.epeter-stats.de/roc-curves-and-ties/}, which was
written by the author of the {\fontseries{b}\selectfont fbroc} (Peter
2016) package, which we will discuss below. Most notably, Hsu and Lieli
(2014) is an extensive discussion of ties, but the paper was not
published.

Although many discuss the properties of ROC and AUC analyses, we will
first show the math and calculations of the AUC with a binary predictor,
which correspond to simple calculations based on sensitivity and
specificity. We then explore commonly-used statistical software for ROC
curve creation and AUC calculation in a variety of packages and
languages. Overall, we believe that AUC calculations alone may be
misleading for binary or categorical predictors depending on the
definition of the AUC. We propose to be explicit when reporting the AUC
in terms of the approach to ties and discuss using step function
interpolation when comparing AUC.

\hypertarget{mathematical-proof-of-auc-for-single-binary-predictor}{%
\section{Mathematical Proof of AUC for Single Binary
Predictor}\label{mathematical-proof-of-auc-for-single-binary-predictor}}

First, we will show how the AUC is defined in terms of probability. This
representation is helpful in discussing the connection between the
stated interpretation of the AUC, the formal definition and calculation
used in software, and how the treatment of ties is crucial when the data
are discrete. Let us assume we have a binary predictor \(X\) and a
binary outcome \(Y\), such that \(X\) and \(Y\) only take the values
\(0\) and \(1\), the number of replicates is not relevant here. Let
\(X_{i}\) and \(Y_{i}\) be the values of subject \(i\).

Fawcett (2006) goes on to state:

\begin{quote}
AUC of a classifier is equivalent to the probability that the classifier
will rank a randomly chosen positive instance higher than a randomly
chosen negative instance.
\end{quote}

In other words, we could discern the definition
\(\text{AUC} = P(X_{i} > X_{j} | Y_{i} = 1, Y_{j} = 0)\) for all
\(i, j\), assuming \((X_{i}, Y_{i}) \independent (X_{j}, Y_{j})\). Note,
the definition here adds no probability when the classifier is tied,
this is a strict inequality. As there are only two outcomes for \(X\),
we can expand this probability using the law of total probability:

\begin{eqnarray*}
P(X_{i} > X_{j} | Y_{i} = 1, Y_{j} = 0) &=& P(X_{i} =1, X_{j} = 0 | Y_{i} = 1, Y_{j} = 0) \\ 
&=& P(X_{i} =1 | X_{j} = 0, Y_{i} = 1, Y_{j} = 0) \\
&\times& P(X_{j} =0 | Y_{i} = 1, Y_{j} = 0) \\ 
&=& P(X_{i} =1 | Y_{i} = 1) P(X_{j} =0 | Y_{j} = 0) 
\end{eqnarray*}

Thus, we see that \(P(X_{i} =1 | Y_{i} = 1)\) is the sensitivity and
\(P(X_{j} =0 | Y_{j} = 0)\) is the specificity, so this reduces to:

\begin{equation}
P(X_{i} > X_{j} | Y_{i} = 1, Y_{j} = 0) = \text{specificity} \times \text{sensitivity} \label{eq:expand}
\end{equation}

Thus, using the definition as
\(P(X_{i} > X_{j} | Y_{i} = 1, Y_{j} = 0)\), the AUC of a binary
predictor is simply the sensitivity times the specificity. We will
define this as the the strict definition of AUC, where ties are not
taken into account and we are using strictly greater than in the
probability and will call this value \(\text{AUC}_{\text{definition}}\).

Let us change the definition of AUC slightly while accounting for ties,
which we call \(\text{AUC}_{\text{w/ties}}\), to: \[
\text{AUC}_{\text{w/ties}} = P(X_{i} > X_{j} | Y_{i} = 1, Y_{j} = 0) + \frac{1}{2} P(X_{i} = X_{j} | Y_{i} = 1, Y_{j} = 0)
\] which corresponds to the common definition of AUC (Fawcett 2006;
Saito and Rehmsmeier 2015; Pepe, Longton, and Janes 2009). This AUC is
the one reported by most software, as we will see below.

\hypertarget{simple-concrete-example}{%
\subsection{Simple Concrete Example}\label{simple-concrete-example}}

To give some intuition of this scenario, we will assume \(X\) and \(Y\)
have the following joint distribution, where \(X\) is along the rows and
\(Y\) is along the columns, as in Table \ref{tab:create_tab}.

\begin{table}[ht]

\caption{\label{tab:create_tab}A simple 2x2 table of a binary predictor (rows) versus a binary outcome (columns)}
\centering
\begin{tabular}{l|r|r}
\hline
  & 0 & 1\\
\hline
0 & 52 & 35\\
\hline
1 & 32 & 50\\
\hline
\end{tabular}
\end{table}

Therefore, the AUC should be equal to
\(\frac{50}{85} \times \frac{52}{84}\), which equals 0.364. This
estimated AUC will be reported throughout the paper, so note the value.

Note, if we reverse the labels, then the sensitivity and the specificity
are estimated by \(1\) minus that measure, or
\(\frac{35}{85} \times \frac{32}{84}\), which is equal to 0.157. Thus,
as this AUC is less than the original labeling, we would choose that
with the original labeling.

If we used the calculation for \(\text{AUC}_{\text{w/ties}}\) we see
that we estimate AUC by
\(\text{AUC}_{\text{definition}} + \frac{1}{2}\left( \frac{50 + 52}{169}\right)\),
which is equal to 0.604. We will show that most software report this AUC
estimate.

\hypertarget{monte-carlo-estimation-of-auc}{%
\subsubsection{Monte Carlo Estimation of
AUC}\label{monte-carlo-estimation-of-auc}}

We can also show that if we use simple Monte Carlo sampling, we can
randomly choose \(X_{i} | Y_{i} = 0\) and \(X_{j} | Y_{j} = 1\). From
these samples, we can estimate these AUC based on the definitions above.
Here, the function \texttt{est.auc} samples \(\ensuremath{10^{6}}\)
random samples from \(X_{i} | Y_{i} = 0\) and \(X_{j} | Y_{j} = 1\),
determines which is greater, or if they are tied, and then calculates
\(\widehat{\text{AUC}}_{\text{definition}}\) and
\(\widehat{\text{AUC}}_{\text{w/ties}}\):

\begin{Shaded}
\begin{Highlighting}[]
\NormalTok{est.auc =}\StringTok{ }\ControlFlowTok{function}\NormalTok{(x, y, }\DataTypeTok{n =} \DecValTok{1000000}\NormalTok{) \{}
\NormalTok{  x1 =}\StringTok{ }\NormalTok{x[y }\OperatorTok{==}\StringTok{ }\DecValTok{1}\NormalTok{] }\CommentTok{# x | y = 1}
\NormalTok{  x0 =}\StringTok{ }\NormalTok{x[y }\OperatorTok{==}\StringTok{ }\DecValTok{0}\NormalTok{] }\CommentTok{# x | y = 0}
\NormalTok{  c1 =}\StringTok{ }\KeywordTok{sample}\NormalTok{(x1, }\DataTypeTok{size =}\NormalTok{ n, }\DataTypeTok{replace =} \OtherTok{TRUE}\NormalTok{)}
\NormalTok{  c0 =}\StringTok{ }\KeywordTok{sample}\NormalTok{(x0, }\DataTypeTok{size =}\NormalTok{ n, }\DataTypeTok{replace =} \OtherTok{TRUE}\NormalTok{)}
\NormalTok{  auc.defn =}\StringTok{ }\KeywordTok{mean}\NormalTok{(c1 }\OperatorTok{>}\StringTok{ }\NormalTok{c0) }\CommentTok{# strictly greater}
\NormalTok{  auc.wties =}\StringTok{ }\NormalTok{auc.defn }\OperatorTok{+}\StringTok{ }\DecValTok{1}\OperatorTok{/}\DecValTok{2} \OperatorTok{*}\StringTok{ }\KeywordTok{mean}\NormalTok{(c1 }\OperatorTok{==}\StringTok{ }\NormalTok{c0) }\CommentTok{# half for ties}
  \KeywordTok{return}\NormalTok{(}\KeywordTok{c}\NormalTok{(}\DataTypeTok{auc.definition =}\NormalTok{ auc.defn,}
           \DataTypeTok{auc.wties =}\NormalTok{ auc.wties))}
\NormalTok{\}}
\NormalTok{sample.estauc =}\StringTok{ }\KeywordTok{est.auc}\NormalTok{(x, y)}
\NormalTok{sample.estauc}
\NormalTok{auc.definition      auc.wties }
      \FloatTok{0.364517}       \FloatTok{0.603929} 
\end{Highlighting}
\end{Shaded}

And thus we see these simulations agree with the values estimated above,
with negligible Monte Carlo error.

\hypertarget{geometric-argument-of-auc}{%
\subsubsection{Geometric Argument of
AUC}\label{geometric-argument-of-auc}}

\begin{figure}[H]
\includegraphics[width=1\linewidth]{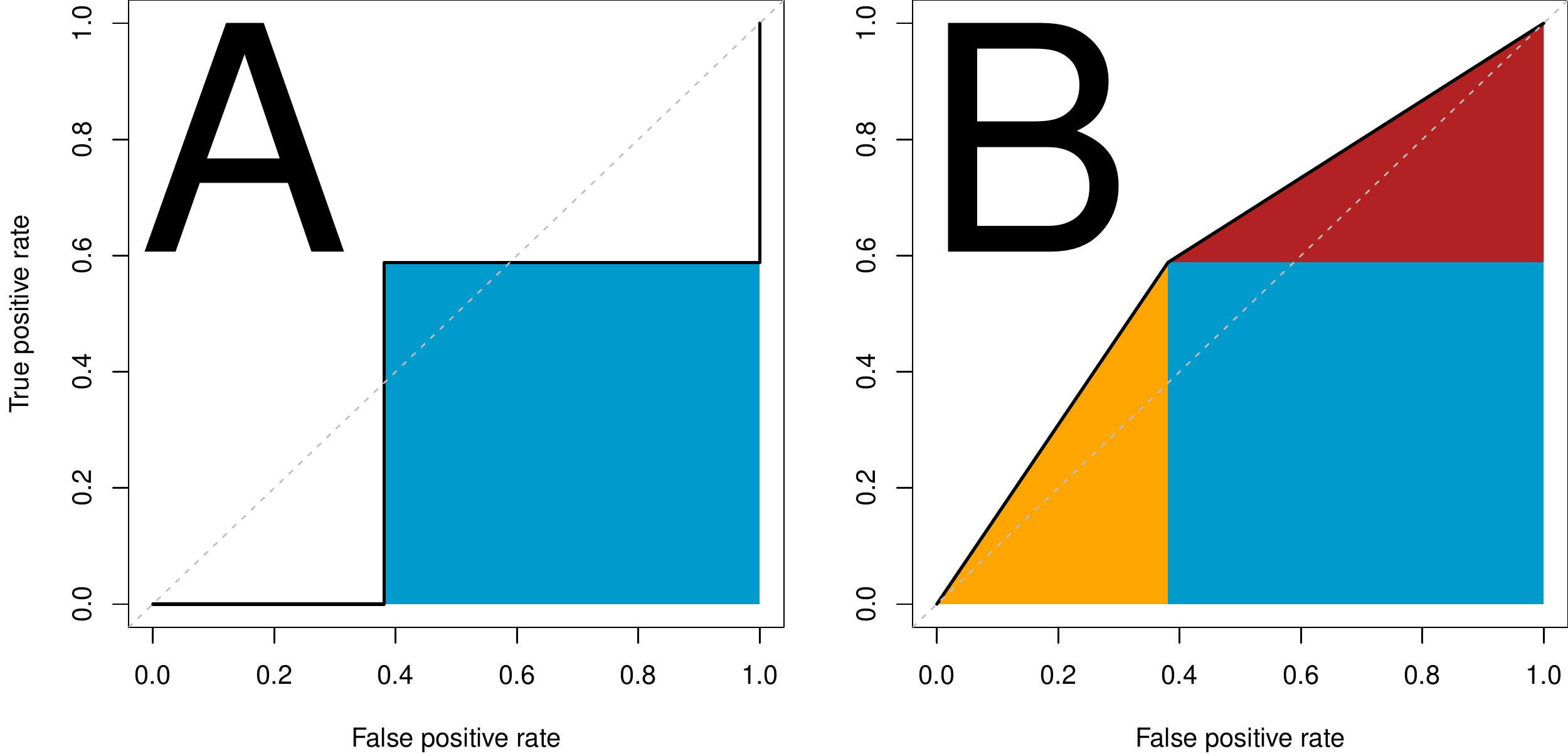} \caption{ROC curve of the data in the simple concrete example.  Here we present a standard ROC curve, with the false positive rate or $1 - \text{specificity}$ on the x-axis and true positive rate or sensitivity on the y-axis.  The dotted line represents the identity. The shaded area in panel represents the AUC for the strict definition.  The additional shaded areas on panel B represent the AUC when accounting for ties.  }\label{fig:main}
\end{figure}

We will present a geometric discussion of the ROC as well. In Figure
\ref{fig:main}, we show the ROC curve for the simple concrete example.
In panel A, we show the point of sensitivity/specificity connected by
the step function, and the associated AUC is represented in the shaded
blue area, representing \(\text{AUC}_{\text{definition}}\). In panel B,
we show the additional shaded areas that are due to ties in orange and
red; all shaded areas represent \(\text{AUC}_{\text{w/ties}}\). We will
show how to calculate these areas from \(P(X_{1} = X_{0})\) in
\(\text{AUC}_{\text{w/ties}}\) such that:


\begin{eqnarray*}
P(X_{i} = X_{j} | Y_{i} = 1, Y_{j} = 0) &=& P(X_{i} = 1, X_{j} = 1 | Y_{i} = 1, Y_{j} = 0) \;+ \\
&& P(X_{i} = 0, X_{j} = 0 | Y_{i} = 1, Y_{j} = 0) \\
&=& P(X_{i} = 1 | Y_{i} = 1) P(X_{j} = 1 | Y_{j} = 0) \;+ \\
&& P(X_{i} = 0 | Y_{i} = 1) P(X_{j} = 0 | Y_{j} = 0) \\
&=& \left(\text{sensitivity} \times (1 - \text{specificity})\right) \;+ \\
&& \left((1- \text{sensitivity}) \times \text{specificity}\right)
\end{eqnarray*}

so that combining this with \eqref{eq:expand} we have:

\begin{align*}
\text{AUC}_{\text{w/ties}} &= \text{specificity} \times \text{sensitivity} \;+ \\
&\hspace{1.3em} \frac{1}{2} \left(\text{sensitivity} \times (1 - \text{specificity})\right) \;+ \\
&\hspace{1.3em} \frac{1}{2} \left((1- \text{sensitivity}) \times \text{specificity}\right) 
\end{align*}

Thus, we can see that geometrically from Figure \ref{fig:main}:

\begin{align*}
\text{AUC}_{\text{w/ties}} &= \includegraphics[width=1.5in,keepaspectratio]{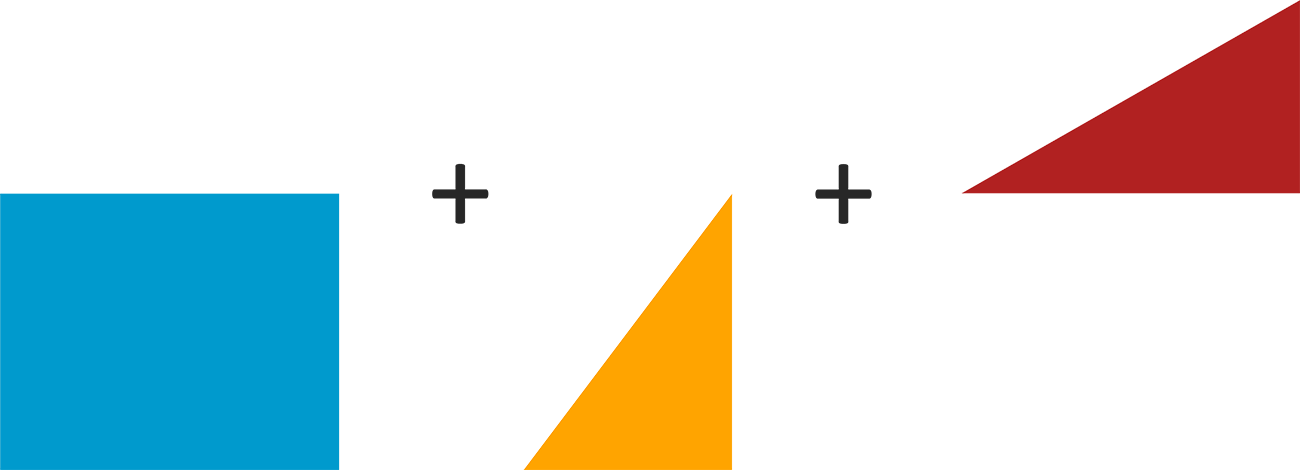}
\end{align*}

where the order of the addition is the same respectively. Note that this
equation reduces further such that:

\begin{align*}
\text{AUC}_{\text{w/ties}} = \frac{1}{2} \left(\text{sensitivity} + \text{specificity}\right).
\end{align*}

Thus, we have shown both \(\text{AUC}_{\text{w/ties}}\) and
\(\text{AUC}_{\text{definition}}\) definition for binary predictors
involve only sensitivity and specificity and can reduce to simple forms.

We will discuss implementations of estimating AUC in software in the
next section. Though we focus on the AUC throughout this paper, many
times ROC analysis is used to find optimal cutoffs for predictors to
give high levels of sensitivity and specificity. The use of the linear
interpolation in the ROC curve gives the false impression that varying
levels of sensitivity and specificity can be achieved by that predictor.
In fact, only the observed sensitivity and specificity can be observed,
other than the trivial cases where sensitivity or specificity is 1. New
samples of the same measurement may give different values, but the
observed measurement can only achieve one point on that curve. Using a
step function interpolation when plotting an ROC curve more clearly
shows this fact.

\hypertarget{auc-calculation-in-statistical-software}{%
\subsection{AUC Calculation in Statistical
Software}\label{auc-calculation-in-statistical-software}}

To determine how these calculations are done in practice, we will
explore the estimated ROC curve and AUC from the implementations in the
following \texttt{R} (R Core Team 2018) packages:
{\fontseries{b}\selectfont ROCR} (Sing et al. 2005),
{\fontseries{b}\selectfont caTools} (Tuszynski 2018),
{\fontseries{b}\selectfont pROC} (Robin et al. 2011), and
{\fontseries{b}\selectfont fbroc} (Peter 2016). We will also show these
agree with the Python implementation in \texttt{sklearn.metrics} from
{\fontseries{b}\selectfont scikit-learn} (Pedregosa et al. 2011), the
Stata functions \texttt{roctab} and \texttt{rocreg} (Bamber 1975;
DeLong, DeLong, and Clarke-Pearson 1988), and the SAS software functions
\texttt{proc\ logistic} with \texttt{roc} and \texttt{roccontrast}. We
note that the majority of these functions all count half the probability
of ties, but differences exist in the calculation of confidence
intervals of AUC and note some inconsistent behavior.

\hypertarget{auc-calculation-current-implementations}{%
\section{AUC Calculation: Current
Implementations}\label{auc-calculation-current-implementations}}

This section will present code and results from commonly-used
implementations of AUC estimation from R, Python, Stata, and SAS
software. We will note agreement with the definitions of AUC above and
any discrepancies. This section is not to be exhaustive, but give
examples how to calculate AUC in these software and show that these
definitions are consistently used in AUC analysis, primarily
\(\widehat{\text{AUC}}_{\text{w/ties}}\).

\hypertarget{r}{%
\subsection{R}\label{r}}

Here we will show the AUC calculation from the common \texttt{R}
packages for ROC analysis. We will show that each report the value
calculated in \(\text{AUC}_{\text{w/ties}}\). The
{\fontseries{b}\selectfont caTools} (Tuszynski 2018) package calculates
AUC using the \texttt{colAUC} function, taking in predictions as
\texttt{x} and the binary ground truth labels as \texttt{y}:

\begin{Shaded}
\begin{Highlighting}[]
\KeywordTok{library}\NormalTok{(caTools)}
\KeywordTok{colAUC}\NormalTok{(x, y)}
\NormalTok{             [,}\DecValTok{1}\NormalTok{]}
\DecValTok{0}\NormalTok{ vs. }\DecValTok{1} \FloatTok{0.6036415}
\end{Highlighting}
\end{Shaded}

which reports \(\text{AUC}_{\text{w/ties}}\).

\begin{figure}
     \centering
     \subfloat[][Stata ROC Plot ]{\includegraphics[width=0.32\linewidth]{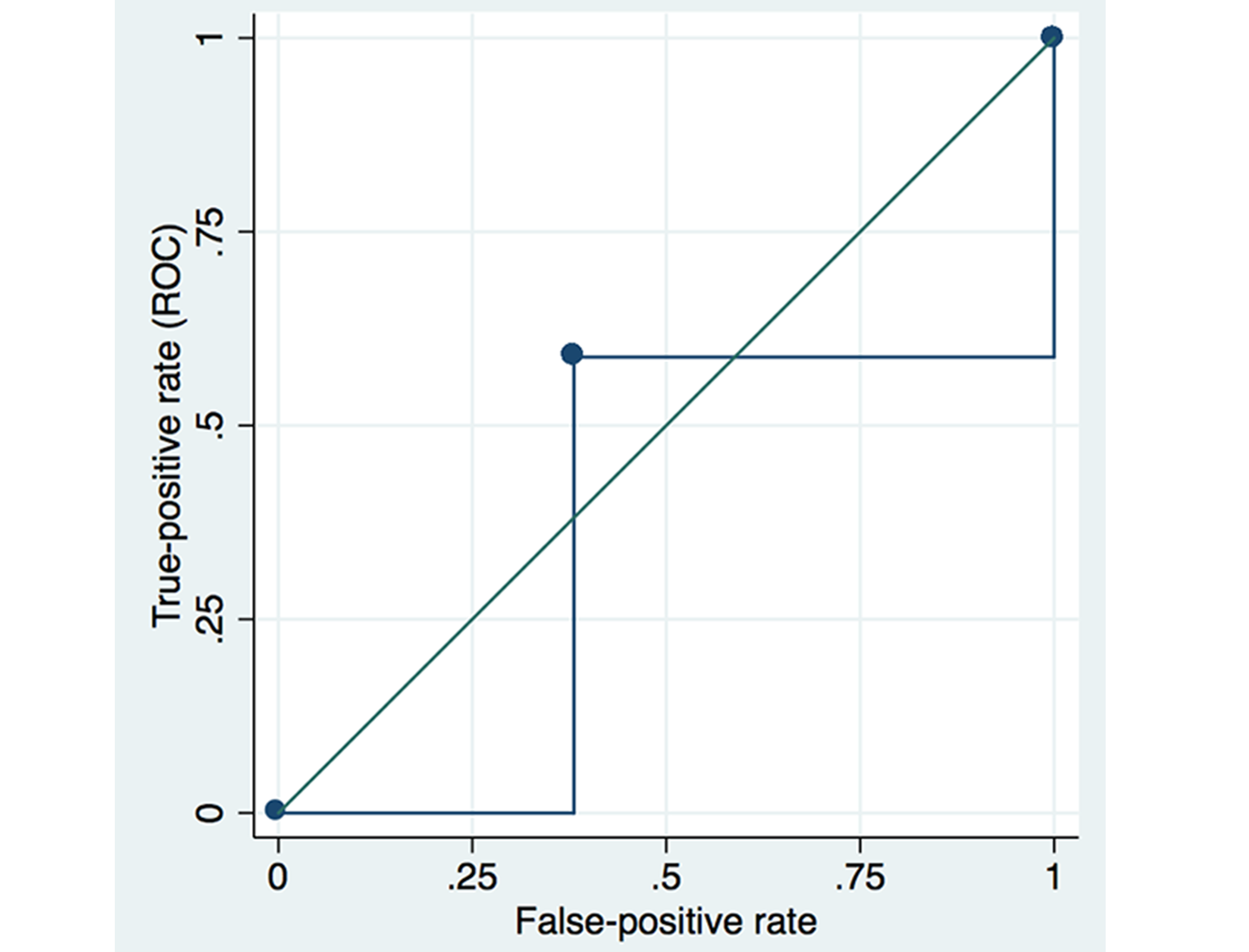}\label{fig:stata}}
     \subfloat[][Python ROC Plot from scikit-learn]{\includegraphics[width=0.32\linewidth]{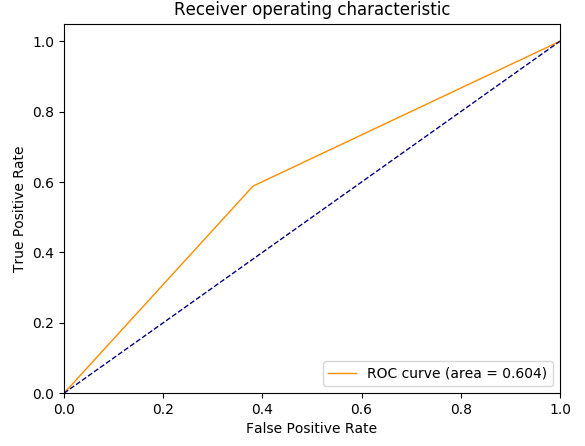}\label{fig:python}} 
     \subfloat[][SAS ROC Plot ]{\includegraphics[width=0.32\linewidth]{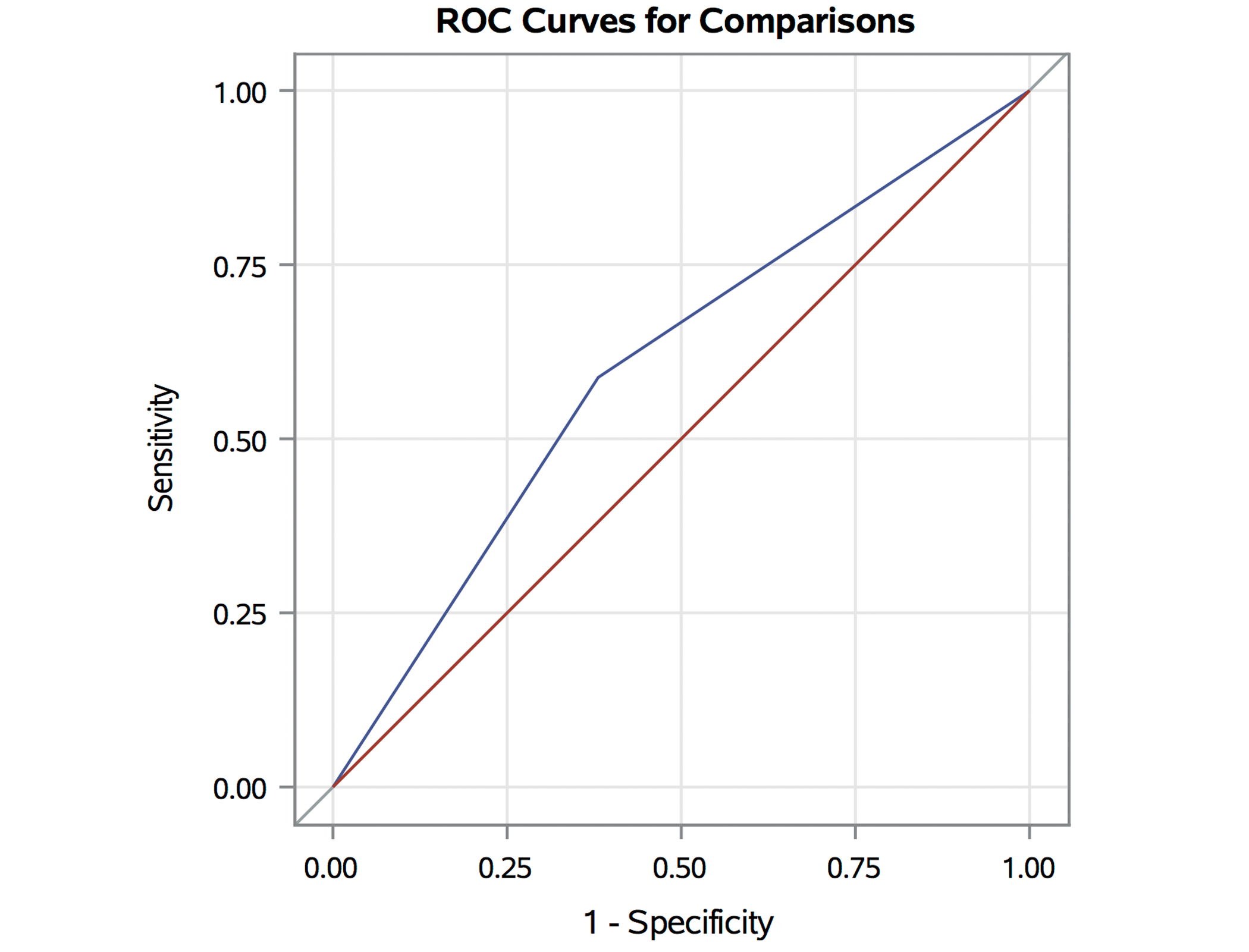}\label{sas}} \\
     \subfloat[][ROCR ROC plot ]{\includegraphics[width=0.32\linewidth]{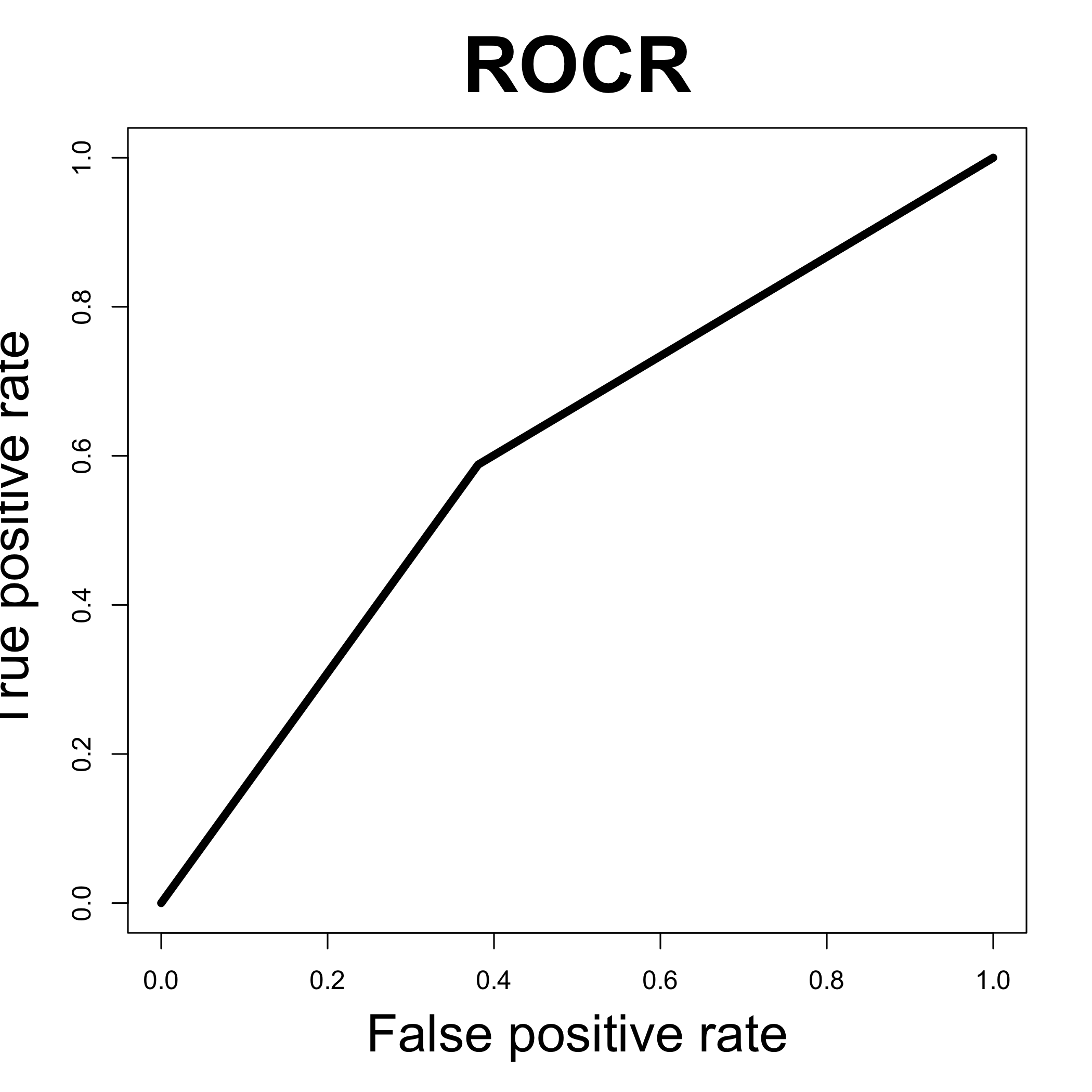}\label{ROCR}}
     \subfloat[][pROC ROC plot ]{\includegraphics[width=0.32\linewidth]{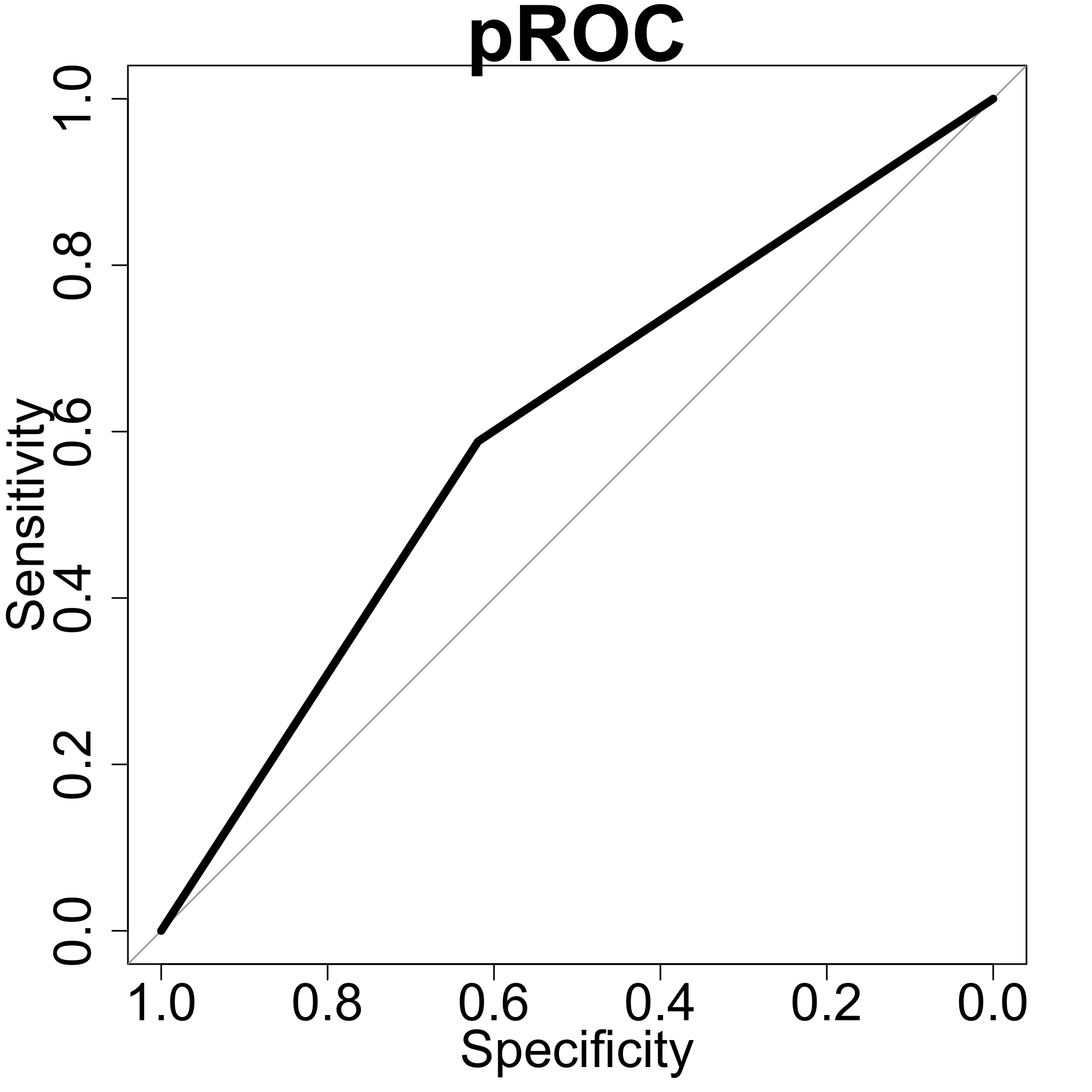}\label{pROC}}
     \subfloat[][fbroc Strategy 2 (default)
]{\includegraphics[width=0.32\linewidth]{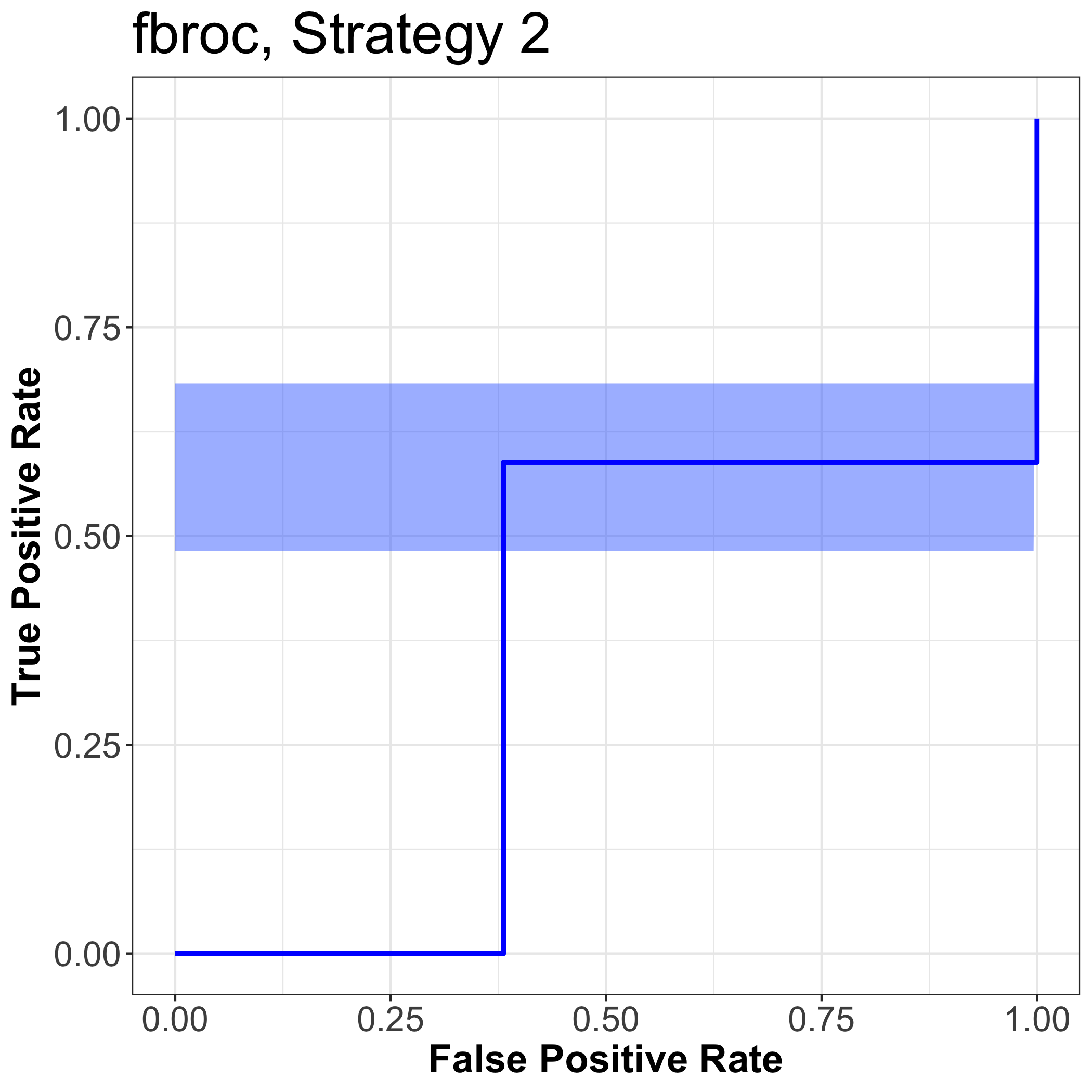}\label{fbroc2}}
     \caption{Comparison of different ROC curves for different  \texttt{R} packages,  \texttt{scikit-learn} from  \texttt{Python},  \texttt{SAS}, and  \texttt{Stata}.  Each line represents the ROC curve, which corresponds to an according area under the curve (AUC).  The blue shading represents the confidence interval for the ROC curve in the {\fontseries{b}\selectfont fbroc} package.  Also, each software represents the curve as the false positive rate versus the true positive rate, though the {\fontseries{b}\selectfont pROC} package calls it sensitivity and specificity (with flipped axes).  Some put the identity line where others do not.  Overall the difference of note as to whether the ROC curve is represented by a step or a linear function. Using the first tie strategy for ties (non-default, not shown) in {\fontseries{b}\selectfont fbroc} gives the same confidence interval but an ROC curve using linear interpolation.}
     \label{fig:rocs}
\end{figure}

In {\fontseries{b}\selectfont ROCR} package (Sing et al. 2005), one must
create a \texttt{prediction} object with the \texttt{prediction}
function, which can calculate a series of measures. AUC is calculated
from a \texttt{performance} function, giving a \texttt{performance}
object, and giving the \texttt{"auc"} measure. We can then extract the
AUC as follows:

\begin{Shaded}
\begin{Highlighting}[]
\KeywordTok{library}\NormalTok{(ROCR)}
\NormalTok{pred =}\StringTok{ }\KeywordTok{prediction}\NormalTok{(x, y)}
\NormalTok{auc.est =}\StringTok{ }\KeywordTok{performance}\NormalTok{(pred, }\StringTok{"auc"}\NormalTok{)}
\NormalTok{auc.est}\OperatorTok{@}\NormalTok{y.values[[}\DecValTok{1}\NormalTok{]]}
\NormalTok{[}\DecValTok{1}\NormalTok{] }\FloatTok{0.6036415}
\end{Highlighting}
\end{Shaded}

which reports \(\text{AUC}_{\text{w/ties}}\). We see this agrees with
the plot from \texttt{ROCR} in Figure \ref{ROCR}.

The {\fontseries{b}\selectfont pROC} (Robin et al. 2011) package
calculates AUC using the \texttt{roc} function:

\begin{Shaded}
\begin{Highlighting}[]
\KeywordTok{library}\NormalTok{(pROC)}
\NormalTok{pROC.roc =}\StringTok{ }\NormalTok{pROC}\OperatorTok{::}\KeywordTok{roc}\NormalTok{(}\DataTypeTok{predictor =}\NormalTok{ x, }\DataTypeTok{response =}\NormalTok{ y)}
\NormalTok{pROC.roc[[}\StringTok{"auc"}\NormalTok{]]}
\NormalTok{Area under the curve}\OperatorTok{:}\StringTok{ }\FloatTok{0.6036}
\end{Highlighting}
\end{Shaded}

which reports \(\text{AUC}_{\text{w/ties}}\) and agrees with the plot
from \texttt{pROC} in Figure \ref{pROC}.

The {\fontseries{b}\selectfont fbroc} package calculates the ROC using
the \texttt{boot.roc} and \texttt{perf} functions. The package has 2
strategies for dealing with ties, which we will create 2 different
objects \texttt{fbroc.default}, using the default strategy (strategy 2),
and alternative strategy (strategy 1, \texttt{fbroc.alternative}):

\begin{Shaded}
\begin{Highlighting}[]
\KeywordTok{library}\NormalTok{(fbroc)}
\NormalTok{fbroc.default =}\StringTok{ }\KeywordTok{boot.roc}\NormalTok{(x, }\KeywordTok{as.logical}\NormalTok{(y), }
                         \DataTypeTok{n.boot =} \DecValTok{1000}\NormalTok{, }\DataTypeTok{tie.strategy =} \DecValTok{2}\NormalTok{)}
\NormalTok{auc.def =}\StringTok{ }\KeywordTok{perf}\NormalTok{(fbroc.default, }\StringTok{"auc"}\NormalTok{)}
\NormalTok{auc.def[[}\StringTok{"Observed.Performance"}\NormalTok{]]}
\NormalTok{[}\DecValTok{1}\NormalTok{] }\FloatTok{0.6036415}
\NormalTok{fbroc.alternative =}\StringTok{ }\KeywordTok{boot.roc}\NormalTok{(x, }\KeywordTok{as.logical}\NormalTok{(y), }
                             \DataTypeTok{n.boot =} \DecValTok{1000}\NormalTok{, }\DataTypeTok{tie.strategy =} \DecValTok{1}\NormalTok{)}
\NormalTok{auc.alt =}\StringTok{ }\KeywordTok{perf}\NormalTok{(fbroc.alternative, }\StringTok{"auc"}\NormalTok{)}
\NormalTok{auc.alt[[}\StringTok{"Observed.Performance"}\NormalTok{]]}
\NormalTok{[}\DecValTok{1}\NormalTok{] }\FloatTok{0.6036415}
\end{Highlighting}
\end{Shaded}

which both report \(\text{AUC}_{\text{w/ties}}\), though the plot from
{\fontseries{b}\selectfont fbroc} in Figure \ref{fbroc2}, which is for
strategy 2, shows a step function, corresponding to
\(\text{AUC}_{\text{definition}}\).

\begin{figure}
     \centering
     \subfloat[][fbroc Strategy 1  ]{\includegraphics[width=0.48\linewidth]{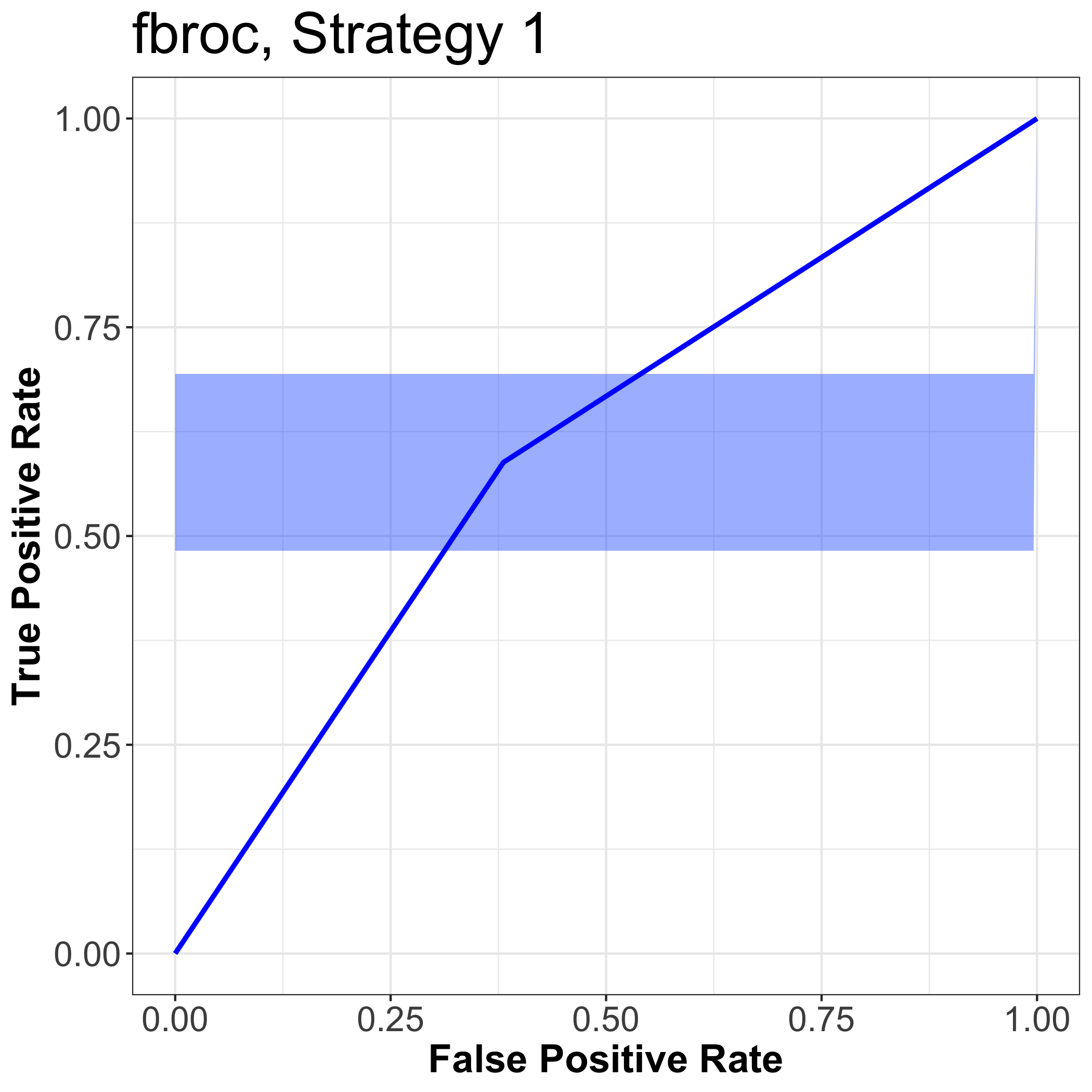}\label{fig:fbroc1}}
     \subfloat[][fbroc Strategy 2 (default)
]{\includegraphics[width=0.48\linewidth]{fbroc2.png}\label{fig:fbroc2}}
     \caption{Comparison of different strategies for ties in the  {\fontseries{b}\selectfont fbroc} package.  The blue shading represents the confidence interval for the ROC curve.  Overall the difference of note as to whether the ROC curve is represented by a step or a linear function. Using the first tie strategy for ties (non-default) in {\fontseries{b}\selectfont fbroc} gives the same confidence interval as the second strategy but an ROC curve using linear interpolation, which may give an inconsistent combination of estimate and confidence interval as {\fontseries{b}\selectfont fbroc} reports the AUC corresponding to the linear interpolation.}
     \label{fig:fbrocs}
\end{figure}

Although the output is the same, these strategies for ties are different
for the plotting for the ROC curve, which we see in Figure
\ref{fig:fbrocs}. The standard error calculation for both strategies use
the second strategy (Fawcett's ``pessimistic'' approach), which is
described in a blog post
(\url{https://www.epeter-stats.de/roc-curves-and-ties/}) and can be seen
in the shaded areas of the panels. Thus, we see that using either tie
strategy results in the same estimate of AUC
(\(\text{AUC}_{\text{w/ties}}\)) and are consistent for tie strategy 1
(Figure \ref{fig:fbroc1}). Using tie strategy 2 results in a plot which
would reflect an AUC of \(\text{AUC}_{\text{definition}}\) (Figure
\ref{fig:fbroc2}), which disagrees with the estimate. This result is
particularly concerning because the plot should agree with the
interpretation of AUC.

\hypertarget{python}{%
\subsection{Python}\label{python}}

In Python, we will use the implementation in \texttt{sklearn.metrics}
from {\fontseries{b}\selectfont scikit-learn} (Pedregosa et al. 2011).
We will use the \texttt{R} package \texttt{reticulate} (Allaire, Ushey,
and Tang 2018), which will provide an Python interface to \texttt{R}.
Here we use the \texttt{roc\_curve} and \texttt{auc} functions from
{\fontseries{b}\selectfont scikit-learn} and output the estimated AUC:

\begin{Shaded}
\begin{Highlighting}[]
\CommentTok{# Adapted from https://qiita.com/bmj0114/items/460424c110a8ce22d945}
\KeywordTok{library}\NormalTok{(reticulate)}
\NormalTok{sk =}\StringTok{ }\KeywordTok{import}\NormalTok{(}\StringTok{"sklearn.metrics"}\NormalTok{)}
\NormalTok{py.roc.curve =}\StringTok{ }\NormalTok{sk}\OperatorTok{$}\KeywordTok{roc_curve}\NormalTok{(}\DataTypeTok{y_score =}\NormalTok{ x, }\DataTypeTok{y_true =}\NormalTok{ y)}
\KeywordTok{names}\NormalTok{(py.roc.curve) =}\StringTok{ }\KeywordTok{c}\NormalTok{(}\StringTok{"fpr"}\NormalTok{, }\StringTok{"tpr"}\NormalTok{, }\StringTok{"thresholds"}\NormalTok{)}
\NormalTok{py.roc.auc =}\StringTok{ }\NormalTok{sk}\OperatorTok{$}\KeywordTok{auc}\NormalTok{(py.roc.curve}\OperatorTok{$}\NormalTok{fpr, py.roc.curve}\OperatorTok{$}\NormalTok{tpr)}
\NormalTok{py.roc.auc}
\NormalTok{[}\DecValTok{1}\NormalTok{] }\FloatTok{0.6036415}
\end{Highlighting}
\end{Shaded}

which reports \(\text{AUC}_{\text{w/ties}}\). Although we have not
exhaustively shown Python reports \(\text{AUC}_{\text{w/ties}}\),
\texttt{scikit-learn} is one of the most popular Python modules for
machine learning and analysis. We can use \texttt{matplotlib} (Hunter
2007) to plot the false positive rate and true positive rate from the
\texttt{py.roc.curve} object, which we see in Figure \ref{fig:python},
which uses a linear interpolation by default and agrees with
\(\text{AUC}_{\text{w/ties}}\).

\hypertarget{sas-software}{%
\subsection{SAS Software}\label{sas-software}}

In SAS software (version 9.4 for Unix) (SAS and Version 2017), let us
assume we have a data set named \texttt{roc} loaded with the
variables/columns of \texttt{x} and \texttt{y} as above. The following
commands will produce the ROC curve in Figure \ref{sas}:

\begin{verbatim}
proc logistic data=roc;
    model y(event='1') = x;
    roc; roccontrast;
    run;      
\end{verbatim}

The resulting output reports \(\text{AUC}_{\text{w/ties}}\), along with
a confidence interval. The calculations can be seen in the SAS User
Guide
(\url{https://support.sas.com/documentation/cdl/en/statug/63033/HTML/default/viewer.htm\#statug_logistic_sect040.htm}),
which includes the addition of the probability of ties.

\hypertarget{stata}{%
\subsection{Stata}\label{stata}}

In Stata (StataCorp, College Station, TX, version 13) (Stata 2013), let
us assume we have a data set with the variables/columns of \texttt{x}
and \texttt{y} as above.

The function \texttt{roctab} is one common way to calculate an AUC:

\begin{verbatim}
roctab x y
                      ROC                    -Asymptotic Normal--
           Obs       Area     Std. Err.      [95% Conf. Interval]
         --------------------------------------------------------
           169     0.6037       0.0379        0.52952     0.67793
\end{verbatim}

which agrees with the calculation based on
\(\text{AUC}_{\text{w/ties}}\) and agrees with the estimates from above.
One can also calculate the AUC using the \texttt{rocreg} function:

\begin{verbatim}
rocreg y x, nodots auc
Bootstrap results                               Number of obs      =       169
                                                Replications       =      1000
Nonparametric ROC estimation
Control standardization: empirical
ROC method             : empirical
Area under the ROC curve
   Status    : y
   Classifier: x
------------------------------------------------------------------------------
             |    Observed               Bootstrap
         AUC |       Coef.       Bias    Std. Err.     [95% Conf. Interval]
-------------+----------------------------------------------------------------
             |    .3641457  -.0004513    .0451334     .2756857   .4526056  (N)
             |                                        .2771778    .452824  (P)
             |                                        .2769474   .4507576 (BC)
------------------------------------------------------------------------------
\end{verbatim}

which agrees with \(\text{AUC}_{\text{definition}}\) and is different
from the output from \texttt{roctab}. The variance of the estimate is
based on a bootstrap estimate, but the point estimate will remain the
same regardless of using the bootstrap or not. This disagreement of
estimates is concerning as the reported estimated AUC may be different
depending on the command used in the estimation.

Using \texttt{rocregplot} after running this estimation, we see can
create an ROC curve, which is shown in Figure \ref{fig:stata}. We see
that the estimated ROC curve coincides with the estimated AUC from
\texttt{rocreg} (\(\text{AUC}_{\text{definition}}\)) and the blue
rectangle in Figure \ref{fig:main}. Thus, \texttt{roctab} is one of the
most common ways in Stata to estimate AUC, but does not agree with the
common way to plot ROC curves.

Thus, we see in Figure \ref{fig:rocs} that all ROC curves are
interpolated with a linear interpolation, which coincides with the
calculation based on \(\text{AUC}_{\text{w/ties}}\), except for the
Stata and {\fontseries{b}\selectfont fbroc} ROC curves, which
interpolates using a step function and coincides with
\(\text{AUC}_{\text{definition}}\). The confidence interval estimate of
the ROC curve for {\fontseries{b}\selectfont fbroc}, which is shaded in
blue in Figure \ref{fbroc2}, corresponds to variability based on
\(\text{AUC}_{\text{definition}}\), but the reported value corresponds
to the ROC curve based on \(\text{AUC}_{\text{w/ties}}\).

Figure \ref{fig:fbrocs} shows that using the different tie strategies
gives a linear (strategy 2, default, panel \subref{fig:fbroc2},
duplicated) or step function/constant (strategy 1, panel
\subref{fig:fbroc1}) interpolation. In each tie strategy, however, the
AUC is \textbf{estimated to be the same}. Therefore, tie strategy 2 may
give an inconsistent combination of AUC estimate and ROC representation
and strategy 1 may give an inconsistent estimation of the variability of
the ROC.

\hypertarget{categorical-predictor-example}{%
\section{Categorical Predictor
Example}\label{categorical-predictor-example}}

Though the main focus of the paper is to demonstrate how using an AUC
directly on a binary predictor can lead to overestimation of predictive
power, we believe this relates to categorical values as well. With
binary predictors, using single summary measures such as specificity and
sensitivity can and should be conveyed for performance measures. With
categorical predictors, however, multiple thresholds are available and
simple one-dimensional summaries are more complicated and ROC curves may
give insight. Let us assume we had a categorical variable, such as one
measured using a 4-point Likert scale, which takes on the following
cross-tabulation with the outcome in Table
\ref{tab:create_cat_tab_output}.

\begin{table}[ht]

\caption{\label{tab:create_cat_tab_output}A example table of a categorical predictor (rows) versus a binary outcome (columns)}
\centering
\begin{tabular}{l|r|r}
\hline
  & 0 & 1\\
\hline
1 & 31 & 21\\
\hline
2 & 21 & 14\\
\hline
3 & 11 & 17\\
\hline
4 & 21 & 33\\
\hline
\end{tabular}
\end{table}

\begin{figure}[H]
\includegraphics[width=1\linewidth]{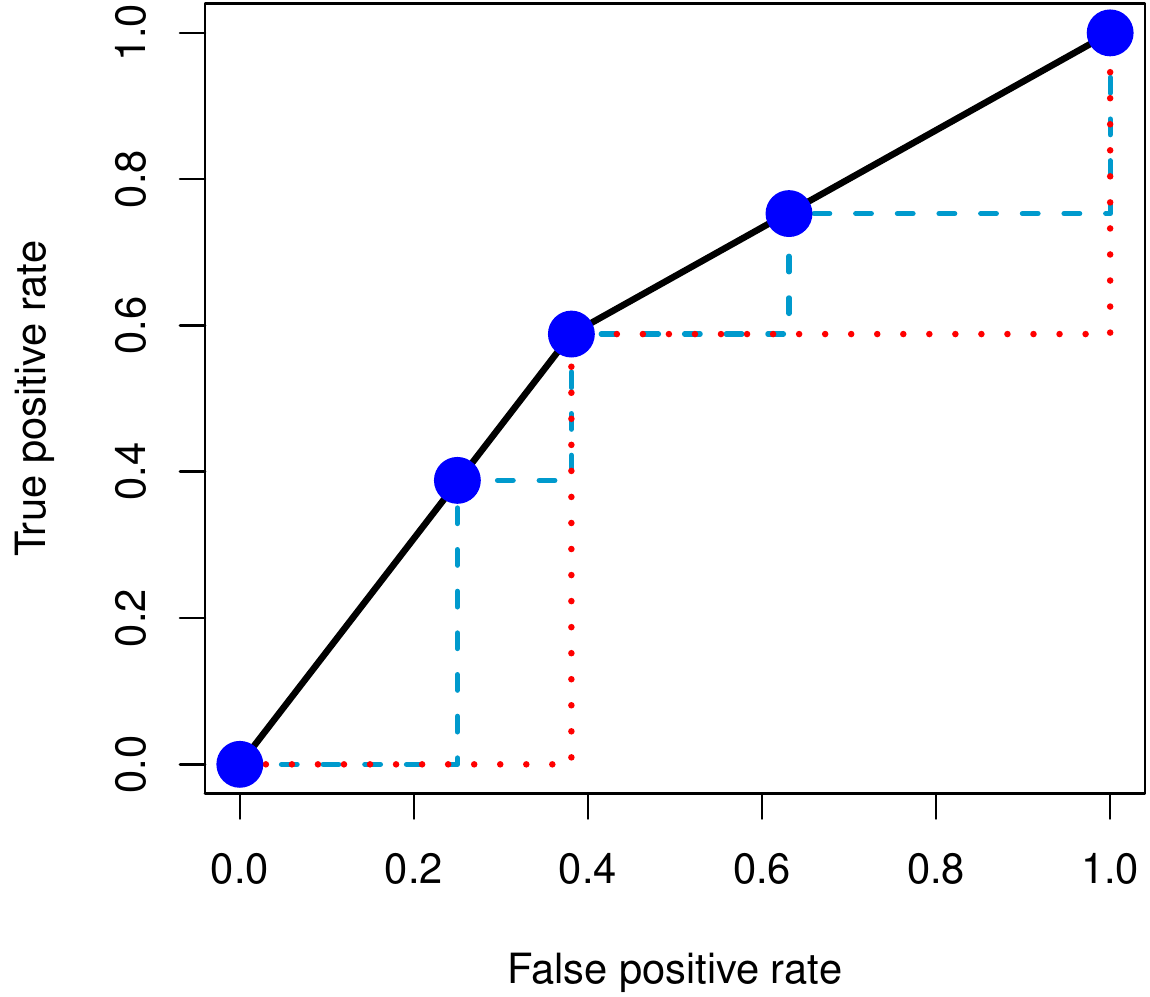} \caption{ROC curve of a 4-level categorical variable compared to the binary predictor. Here we present the ROC curve of a categorical predictor (blue points) compared to that of the binary predictor (black line).  We see that the ROC curve is identical if the linear inerpolation is used (accounting for ties).  The red (dotted) and blue (dashed) lines show the ROC of the binary and categorical predictor, respectively, using the pessimistic approach.  We believe this demonstrates that although there is more gradation in the categorical variable, using the standard approach provides the same AUC, though we believe these variables have different levels of information as the binary predictor cannot obtain values other than the 2 categories.  }\label{fig:maincat}
\end{figure}

Note, the number of records is the same as in the binary predictor case.
In Figure \ref{fig:maincat}, we overlay the ROC curves from this
predictor and the binary predictor (black line), showing they are the
same, with points on the ROC curve from the categorical predictor in
blue. We also show the ROC curve with lines using the pessimistic
approach for the categorical predictor (blue, dashed) and the binary
predictor (red, dotted).

We see that using the the linear interpolation in Figure
\ref{fig:maincat}, the AUC for the categorical predictor and the binary
predictor would be nearly identical. The AUC is not exact due to the
fact that the cells in the table must be integers. This result is
counterintuitive as the categorical variable can take on a number of
values with varying sensitivities and specificities, whereas the binary
predictor cannot. With more samples, we could extend this example to a
variable which had hundreds of values like a continuous predictor, but
give the same ROC shape and identical AUC when using linear
interpolation. We show an example in the supplemental material. Using
the pessimistic approach (red dotted and blue dashed lines in Figure
\ref{fig:maincat}) for the categorical and binary predictors, we see
that the AUC would be different in this estimation.

\hypertarget{conclusion}{%
\section{Conclusion}\label{conclusion}}

We have shown how the ROC curve is plotted and AUC is estimated in
common statistical software when using a univariate binary predictor.
There are inconsistencies across software platforms, such as \texttt{R}
and Stata, and even within some packages, such as
{\fontseries{b}\selectfont fbroc}. We believe these calculations may not
reflect the discreteness of the data. We agree that using a binary
predictor in an ROC analysis may not be appropriate, but we note that
researchers and users may still perform this analysis.

We believe the critiques depend partially of the nature of the
predictor. Some predictors are fundamentally discrete or categorical,
such as the number of different alleles at a gene or a questionnaire
using Likert scales. Others are continuous but empirically discrete
either by rounding or a small set of unique values. For predictors that
are not fundamentally discrete, we believe that linear interpolation
would be reasonable if unobserved values in between those observed are
theoretically possible.

Otherwise, we believe using the step function interpolation and not
counting ties would be more appropriate. We believe additional options
for different calculations accounting for ties should be possible or
warnings for discrete data may be presented to the user. We hope that
indicating how ties are handled would become more common, especially for
discrete data in practice. Using different methods for ties poses
different issues, such as AUC values that are below \(0.5\) and some
tests may not have the same theoretical properties or connections to
Wilcoxon rank-sum tests. Though these new issues arise, we believe the
current methodology has the potential for misleading users.

All code required to generate this paper is located at
\url{https://github.com/muschellij2/binroc}.

\hypertarget{references}{%
\section*{References}\label{references}}
\addcontentsline{toc}{section}{References}

\hypertarget{refs}{}
\leavevmode\hypertarget{ref-reticulate}{}%
Allaire, JJ, Kevin Ushey, and Yuan Tang. 2018. \emph{reticulate:
Interface to 'Python'}. \url{https://github.com/rstudio/reticulate}.

\leavevmode\hypertarget{ref-bamber1975area}{}%
Bamber, Donald. 1975. ``The Area Above the Ordinal Dominance Graph and
the Area Below the Receiver Operating Characteristic Graph.''
\emph{Journal of Mathematical Psychology} 12 (4): 387--415.

\leavevmode\hypertarget{ref-blumberg2016technology}{}%
Blumberg, Dana M, De MoraesCarlos Gustavo, Jeffrey M Liebmann, Reena
Garg, Cynthia Chen, Alex Theventhiran, and Donald C Hood. 2016.
``Technology and the Glaucoma Suspect.'' \emph{Investigative
Ophthalmology \& Visual Science} 57 (9): OCT80--OCT85.

\leavevmode\hypertarget{ref-budwega2016factors}{}%
Budwega, Joris, Till Sprengerb, De Vere-TyndalldAnthony, Anne
Hagenkordd, Christoph Stippichd, and Christoph T Bergera. 2016.
``Factors Associated with Significant MRI Findings in Medical Walk-in
Patients with Acute Headache.'' \emph{Swiss Med Wkly} 146: w14349.

\leavevmode\hypertarget{ref-delong}{}%
DeLong, Elizabeth R, David M DeLong, and Daniel L Clarke-Pearson. 1988.
``Comparing the Areas Under Two or More Correlated Receiver Operating
Characteristic Curves: A Nonparametric Approach.'' \emph{Biometrics},
837--45.

\leavevmode\hypertarget{ref-jama}{}%
E, Maverakis, Ma C, Shinkai K, and et al. 2018. ``Diagnostic Criteria of
Ulcerative Pyoderma Gangrenosum: A Delphi Consensus of International
Experts.'' \emph{JAMA Dermatology} 154 (4): 461--66.
\url{https://doi.org/10.1001/jamadermatol.2017.5980}.

\leavevmode\hypertarget{ref-fawcett2006introduction}{}%
Fawcett, Tom. 2006. ``An Introduction to Roc Analysis.'' \emph{Pattern
Recognition Letters} 27 (8): 861--74.

\leavevmode\hypertarget{ref-glaveckaite2011value}{}%
Glaveckaite, Sigita, Nomeda Valeviciene, Darius Palionis, Viktor
Skorniakov, Jelena Celutkiene, Algirdas Tamosiunas, Giedrius Uzdavinys,
and Aleksandras Laucevicius. 2011. ``Value of Scar Imaging and Inotropic
Reserve Combination for the Prediction of Segmental and Global Left
Ventricular Functional Recovery After Revascularisation.'' \emph{Journal
of Cardiovascular Magnetic Resonance} 13 (1): 35.

\leavevmode\hypertarget{ref-hanley1982meaning}{}%
Hanley, James A, and Barbara J McNeil. 1982. ``The Meaning and Use of
the Area Under a Receiver Operating Characteristic (ROC) Curve.''
\emph{Radiology} 143 (1): 29--36.

\leavevmode\hypertarget{ref-hsu2014inference}{}%
Hsu, Yu-Chin, and R Lieli. 2014. ``Inference for ROC Curves Based on
Estimated Predictive Indices: A Note on Testing AUC = 0.5.''
\emph{Unpublished Manuscript}.

\leavevmode\hypertarget{ref-matplotlib}{}%
Hunter, J. D. 2007. ``Matplotlib: A 2D Graphics Environment.''
\emph{Computing in Science \& Engineering} 9 (3): 90--95.
\url{https://doi.org/10.1109/MCSE.2007.55}.

\leavevmode\hypertarget{ref-kushnir2018degree}{}%
Kushnir, Vitaly A, Sarah K Darmon, David H Barad, and Norbert Gleicher.
2018. ``Degree of Mosaicism in Trophectoderm Does Not Predict Pregnancy
Potential: A Corrected Analysis of Pregnancy Outcomes Following Transfer
of Mosaic Embryos.'' \emph{Reproductive Biology and Endocrinology} 16
(1): 6.

\leavevmode\hypertarget{ref-mwipatayi2016durability}{}%
Mwipatayi, Bibombe P, Surabhi Sharma, Ali Daneshmand, Shannon D Thomas,
Vikram Vijayan, Nishath Altaf, Marek Garbowski, et al. 2016.
``Durability of the Balloon-Expandable Covered Versus Bare-Metal Stents
in the Covered Versus Balloon Expandable Stent Trial (COBEST) for the
Treatment of Aortoiliac Occlusive Disease.'' \emph{Journal of Vascular
Surgery} 64 (1): 83--94.

\leavevmode\hypertarget{ref-scikitlearn}{}%
Pedregosa, F., G. Varoquaux, A. Gramfort, V. Michel, B. Thirion, O.
Grisel, M. Blondel, et al. 2011. ``Scikit-Learn: Machine Learning in
Python.'' \emph{Journal of Machine Learning Research} 12: 2825--30.

\leavevmode\hypertarget{ref-pepe2009estimation}{}%
Pepe, Margaret, Gary Longton, and Holly Janes. 2009. ``Estimation and
Comparison of Receiver Operating Characteristic Curves.'' \emph{The
Stata Journal} 9 (1): 1.

\leavevmode\hypertarget{ref-fbroc}{}%
Peter, Erik. 2016. \emph{Fbroc: Fast Algorithms to Bootstrap Receiver
Operating Characteristics Curves}.
\url{https://CRAN.R-project.org/package=fbroc}.

\leavevmode\hypertarget{ref-rcore}{}%
R Core Team. 2018. \emph{R: A Language and Environment for Statistical
Computing}. Vienna, Austria: R Foundation for Statistical Computing.
\url{https://www.R-project.org/}.

\leavevmode\hypertarget{ref-pROC}{}%
Robin, Xavier, Natacha Turck, Alexandre Hainard, Natalia Tiberti,
Frédérique Lisacek, Jean-Charles Sanchez, and Markus Müller. 2011.
``pROC: An Open-Source Package for R and S+ to Analyze and Compare ROC
Curves.'' \emph{BMC Bioinformatics} 12: 77.

\leavevmode\hypertarget{ref-saito2015precision}{}%
Saito, Takaya, and Marc Rehmsmeier. 2015. ``The Precision-Recall Plot Is
More Informative Than the ROC Plot When Evaluating Binary Classifiers on
Imbalanced Datasets.'' \emph{PloS One} 10 (3): e0118432.

\leavevmode\hypertarget{ref-sas}{}%
SAS, SAS, and STAT Version. 2017. ``9.4 {[}Computer Program{]}.''
\emph{Cary, NC: SAS Institute}.

\leavevmode\hypertarget{ref-shterev2018bayesian}{}%
Shterev, Ivo D, David B Dunson, Cliburn Chan, and Gregory D Sempowski.
2018. ``Bayesian Multi-Plate High-Throughput Screening of Compounds.''
\emph{Scientific Reports} 8 (1): 9551.

\leavevmode\hypertarget{ref-ROCR}{}%
Sing, T., O. Sander, N. Beerenwinkel, and T. Lengauer. 2005. ``ROCR:
Visualizing Classifier Performance in R.'' \emph{Bioinformatics} 21
(20): 7881. \url{http://rocr.bioinf.mpi-sb.mpg.de}.

\leavevmode\hypertarget{ref-snarr2017parasternal}{}%
Snarr, Brian S, Michael Y Liu, Jeremy C Zuckerberg, Christine B
Falkensammer, Sumekala Nadaraj, Danielle Burstein, Deborah Ho, et al.
2017. ``The Parasternal Short-Axis View Improves Diagnostic Accuracy for
Inferior Sinus Venosus Type of Atrial Septal Defects by Transthoracic
Echocardiography.'' \emph{Journal of the American Society of
Echocardiography} 30 (3): 209--15.

\leavevmode\hypertarget{ref-stata}{}%
Stata, Statt. 2013. ``Release 13. Statistical Software.''
\emph{StataCorp LP, College Station, TX}.

\leavevmode\hypertarget{ref-caTools}{}%
Tuszynski, Jarek. 2018. \emph{caTools: Tools: Moving Window Statistics,
GIF, Base64, ROC AUC, Etc.}
\url{https://CRAN.R-project.org/package=caTools}.

\leavevmode\hypertarget{ref-jama2}{}%
TV, Litvin, Bresnick GH, Cuadros JA, Selvin S, Kanai K, and Ozawa GY.
2017. ``A Revised Approach for the Detection of Sight-Threatening
Diabetic Macular Edema.'' \emph{JAMA Ophthalmology} 135 (1): 62--68.
\url{https://doi.org/10.1001/jamaophthalmol.2016.4772}.

\leavevmode\hypertarget{ref-veltri2018deep}{}%
Veltri, Daniel, Uday Kamath, and Amarda Shehu. 2018. ``Deep Learning
Improves Antimicrobial Peptide Recognition.'' \emph{Bioinformatics} 1:
8.

\leavevmode\hypertarget{ref-xiong2018comparison}{}%
Xiong, Xin, Qi Li, Wen-Song Yang, Xiao Wei, Xi Hu, Xing-Chen Wang, Dan
Zhu, Rui Li, Du Cao, and Peng Xie. 2018. ``Comparison of Swirl Sign and
Black Hole Sign in Predicting Early Hematoma Growth in Patients with
Spontaneous Intracerebral Hemorrhage.'' \emph{Medical Science Monitor:
International Medical Journal of Experimental and Clinical Research} 24:
567.

\bibliographystyle{spbasic}
\bibliography{binroc.bib}

\end{document}